# Systematic Classification of Studies Investigating Social Media Conversations about Long COVID Using a Novel Zero-Shot Transformer Framework

Nirmalya Thakur, Niven Francis Da Guia Fernandes, and Madje Tobi Marc'Avent Tchona

Department of Electrical Engineering and Computer Science, South Dakota School of Mines and Technology, Rapid City, SD 57701, USA
`nirmalya.thakur@sdsmt.edu`
`{niven.fernandes, marc.tchona}@mines.sdsmt.edu`

**Abstract.** Long COVID continues to challenge public health by affecting a considerable number of individuals who have recovered from acute SARS-CoV-2 infection yet endure prolonged and often debilitating symptoms. Social media has emerged as a vital resource for those seeking real-time information, peer support, and validating their health concerns related to Long COVID. This paper examines recent works focusing on mining, analyzing, and interpreting user-generated content on social media platforms to capture the broader discourse on persistent post-COVID conditions. A novel transformer-based zero-shot learning approach serves as the foundation for classifying research papers in this area into four primary categories: Clinical or Symptom Characterization, Advanced NLP or Computational Methods, Policy, Advocacy or Public Health Communication, and Online Communities and Social Support. This methodology achieved an average confidence of 0.7788, with the minimum and maximum confidence being 0.1566 and 0.9928, respectively. This model showcases the ability of advanced language models to categorize research papers without any training data or predefined classification labels, thus enabling a more rapid and scalable assessment of existing literature. This paper also highlights the multifaceted nature of Long COVID research by demonstrating how advanced computational techniques applied to social media conversations can reveal deeper insights into the experiences, symptoms, and narratives of individuals affected by Long COVID.

## 1 Introduction

In December 2019, an outbreak of coronavirus disease 2019 (COVID-19), caused by the severe acute respiratory syndrome coronavirus 2 (SARS-CoV-2), began in China. [1,2]. Even though SARS-CoV-2 is similar in origin to SARS-CoV and MERS-CoV, it has affected public health globally at a much greater scale than any prior coronavirus outbreaks [3]. Early containment efforts, including measures by the Chinese government, did not prevent the disease from rapidly crossing regional and international boundaries [4], leading the World Health Organization (WHO) to declare COVID-19 a global pandemic on March 11, 2021 [5]. According to the WHO, confirmed cases were 776,841,264 worldwide, with 7,075,468 reported deaths as of November 10, 2024 [6]. Although many individuals recover from the acute infection caused by SARS-CoV-2, a significant subset experiences symptoms that remain or appear after what might have been presumed clinical recovery. This phenomenon, known as Long COVID, has been described since the earliest days of the pandemic to include persistent or emerging physical and psychological challenges [7-9]. As per [10], *"Long COVID is defined as a chronic condition that occurs after SARS-CoV-2 infection and is present for at least 3 months. Long COVID includes a wide range of symptoms or conditions that may improve, worsen, or be ongoing"*.

Individuals who experience Long COVID commonly face a broad set of symptoms that may disrupt daily routines and overall well-being. Frequently reported symptoms of Long COVID include shortness of breath, cough, persistent fatigue, post-exertional malaise, difficulty concentrating, memory changes, recurring headache, lightheadedness, fast heart rate, sleep disturbance, problems with taste or smell, bloating, constipation, and diarrhea [11-15]. Such symptoms may persist for three months beyond the initial SARS-CoV-2 infection or even exceed a year [16]. Although some individuals gradually improve, others experience lingering or fluctuating complications that may profoundly affect their physical, psychological, and social health [17,18]. Clinicians commonly reference established guidelines when managing symptoms of Long COVID and any accompanying conditions, such as diabetes, high blood pressure, or POTS, to reduce future complications and enhance the patient's quality of life [19]. Many patients benefit from a combination of therapies - ranging from medications targeting pain or sleep challenges to physical or occupational rehabilitation - along with psychological support to manage both the physical and emotional aspects of Long COVID [20,21]. In addition to this, even though certain medications such as paracetamol or NSAIDs appear to help with specific Long COVID symptoms like fever, there is still no standardized treatment to address the entire spectrum of Long COVID symptoms [22-25].

Social media has been a critical venue for public discussion of COVID-19 since its initial cases in December 2019, evolving into a resource for people seeking real-time information and community support [26-31]. As this pandemic advanced, platforms such as TikTok [32,33], Twitch [34,35], WeChat [36,37], Instagram [38,39], Facebook [40,41], YouTube [42,43], Reddit [44,45], LinkedIn [46,47], X (formerly Twitter) [48,49], Clubhouse [50,51], Discord [52,53], and Snapchat [54,55], became pivotal for gathering firsthand insights into ongoing patient experiences. Traditional methods like surveys and interviews can be constrained by time and location, whereas social media allows continuous, unfiltered accounts of Long COVID symptom experiences and daily struggles. Individuals suffering from Long COVID can document their symptoms, exchange practical advice, and discuss personal setbacks or milestones on social media, leading to the generation of Big Data that researchers from different disciplines and healthcare professionals may analyze to identify evolving patterns. As Long COVID presents multifaceted medical, social, and emotional issues, there has been growing interest in leveraging online platforms to study it from multiple angles. Social media channels facilitate global conversations that can reveal differences in experiences related to healthcare access, post-infection complications, or even public awareness of the severity of a health-related condition [56-58]. Over time, these virtual spaces have also fostered advocacy and grassroots efforts. Hashtags like #LongCOVID [59] have given patients and advocates an active role in discussing everything from specialized clinics to mental health support [60]. Observations of this activity underscore how large-scale social media data can shape public health discourse [61] and even influence policies [62] addressing health-related conditions that are often misunderstood or underdiagnosed.

A paper that categorizes the existing work in this domain will further our understanding of ongoing research related to the mining and analysis of Long COVID-related content on social media. Currently, studies vary widely in scope, encompassing approaches such as sentiment analysis, topic modeling, and network analysis. Each of these research areas focuses on different facets of how individuals report, discuss, and cope with prolonged COVID-19 symptoms online. Without a unifying framework, it becomes difficult to understand where the field has robust evidence, where knowledge gaps persist, and which topics support further investigation. In this paper, we go beyond the bounds of traditional literature reviews by proposing a zero-shot classification pipeline. Instead of relying on a manually annotated dataset, we use a transformer-based model that has been trained on extensive, domain-specific data. This model dynamically classifies each paper into one of the following four categories: Clinical or Symptom Characterization, Advanced NLP or Computational Methods, Policy and Advocacy, and Online Communities and Social Support. These categories were chosen for two primary reasons. First, they reflect the multiple dimensions of Long COVID research - from clinical symptom progression to emerging computational analytics and from large-scale public health initiatives to the online communities and support networks that have formed around the topic. Second, they offer a balance between breadth and specificity: each category captures a distinct research focus, yet together, they accommodate the wide range of investigations currently taking place in this rapidly evolving field. Although these studies are classified into distinct categories in this paper, many works address multiple facets of Long COVID research, rendering any classification flexible rather than absolute. For instance, a paper listed under "Online Communities and Social Support" may also perform a detailed sentiment analysis that aligns with "Advanced NLP or Computational Methods". Such overlaps arise naturally in interdisciplinary research, especially when varied computational methods - like sentiment analysis, topic modeling, and network analysis - are applied to the extensive social media discussions surrounding patient experiences, advocacy efforts, and policy implications. The four broad areas presented here are an organizational guide, highlighting a primary thematic focus without dismissing other significant aspects of each study.

The rest of this paper is organized as follows. First, the methodology is presented, describing the search strategy, inclusion criteria, and the steps taken to develop the proposed zero-shot classification model. Then, the results of this classification pro-cess are presented and discussed. The interdisciplinary applications of this approach are then outlined. We conclude by discussing how these findings inform current research trends and by suggesting directions to extend or refine this approach.

## 2      Methodology

This section is divided into two parts. Section 2.1 presents the methodology that was followed to collect research works that focused on the mining and analysis of Long COVID-related content on social media. The design and working of the proposed zero-shot classification model is presented in Section 2.2.

### 2.1    Mining Recent Works that Focused on the Analysis of Long COVID-related Content on Social Media

A broad literature search was carried out across multiple scholarly databases, including PubMed, Scopus, Web of Science, and Google Scholar, to identify studies focused on mining and analyzing the public discourse about Long COVID on social media. This search aimed to capture Long COVID-related research across diverse fields, such as computer science, health sciences, and social sciences. No papers published before 2020 were included, as the COVID-19 outbreak began in December 2019. The search terms used included - "Long COVID", "post-COVID", and "chronic COVID" - as well as keywords indicative of social media use. In addition, terms like "sentiment analysis", "topic modeling", and "network analysis" were included to ensure the retrieval of studies that used computational or statistical methods to examine the public discourse on social media [63]. By integrating health-related terminology with references to social media platforms and relevant analytical techniques, the search strategy was designed to capture the full breadth of scholarly work investigating the ongoing experiences with Long COVID, as expressed on social media.

Studies were selected for inclusion if they used social media data to investigate any aspect of Long COVID. The main inclusion criteria were that the articles utilized a recognized research methodology - whether qualitative, quantitative, or mixed methods and analyzed data gathered primarily from social media platforms. The selection process also considered whether the authors had sufficiently detailed the nature of their quantitative or qualitative approach. Moreover, ethical practices regarding user data, such as anonymization or compliance with platform terms of service, were taken into account to ensure that privacy concerns were handled responsibly. Studies that just mentioned Long COVID were excluded. Research works such as editorials, letters to the editor, or general news articles, which usually lack methodological details, were excluded. If the essential aspects of a paper were missing - for example, neglecting to report how data were collected - those were also removed from consideration. This approach aimed to retain a set of methodologically sound articles that offered substantive insights into the public discourse about Long COVID on social media platforms.

Upon applying these inclusion and exclusion criteria, a total of 39 papers [67-105] were selected for this study. These works represented a range of methods, including sentiment analysis, qualitative content analysis, topic modeling, and network analysis, and they addressed multiple social media platforms. Thereafter, a transformer-based zero-shot classification model [64-66] was developed to classify these papers into one out of the four thematic categories described below. Although a few papers fit under multiple themes, each was placed wherever its primary emphasis appeared strongest, determined by the model based on confidence values.

(i) Clinical or Symptom Characterization ("Symptom Characterization"): Research that primarily aims to identify or quantify the variety of Long COVID symptoms, from social media data. These studies may include statistical analysis but do not perform extensive sentiment or topic modeling. Their main motivation is to collect clinical or epidemiological insights from user posts.

(ii) Advanced NLP or Computational Methods ("NLP and Modeling"): Studies that specifically emphasize methods like deep transformer networks, topic modeling, sentiment analysis, and other advanced computational approaches. This goes beyond a simple symptom count; it highlights a methods-heavy lens on analyzing social media data.

(iii) Policy, Advocacy, or Public Health Communication ("Policy and Advocacy"): Papers exploring how organizations, governments, or communities develop health communications, handle policy issues, and communicate guidelines.

(iv) Online Communities and Social Support ("Community and Support"): Studies focusing on how individuals find emotional or community support on social media, the way they exchange personal stories, or how group dynamics form around shared experiences. The main emphasis is on the psychosocial aspect and the support social media platforms provide.

In selecting these four categories - Symptom Characterization, NLP and Modeling, Policy and Advocacy, and Community and Support - the primary goal was to reflect distinct yet interrelated dimensions of Long COVID research on social media [106,107]. By grouping each study according to its main focus using the model proposed in this paper, we can more effectively convey how different researchers approach the same foundational subject - Long COVID - from multiple angles. The four thematic categories thus offer both conceptual breadth and analytical depth, ensuring that key aspects of this evolving research area are captured in a way that is both systematic and adaptable. Thereafter, this transformer-based zero-shot classification model was developed, which was set up to assign the 39 papers to these categories. This model did not require any training data or predefined classification labels. The process by which this model was developed is described in Section 2.2.

## 2.2     Design of the Proposed Zero-Shot Classification Model

This section presents the zero-shot classification methodology that integrates advanced embedding architectures and a multi-hypothesis alignment mechanism. This model analyzed the titles and abstracts of the selected research papers, assigning each paper to a single thematic category $c_k$ (where k ∈ [N] and N is the total number of categories). An overview of how this model works is shown in Figure 1.

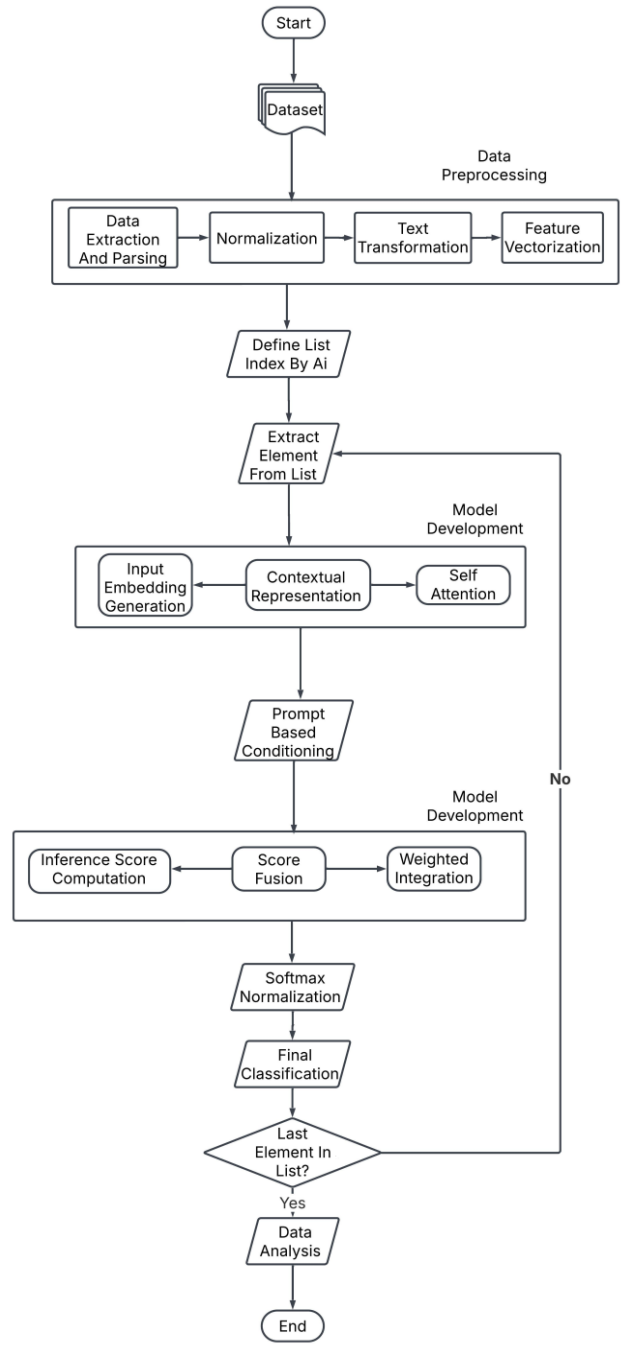

Figure 1. An overview of the working of the zero-shot classification model

A detailed description of this zero-shot classification approach is explained in a series of steps, as shown in Equations (1)–(14). Here, unless noted otherwise, vectors are column vectors, M denotes a matrix, and superscript ⊤ denotes matrix or vector transpose. We define the set of candidate categories as follows:

$$C \triangleq \{c_k \mid c_k \in \Sigma^*, k \in [N]\} \quad (1)$$

where $\Sigma^*$ represents the alphabet of possible label strings (i.e., textual descriptions of categories), and N is the total number of categories in our system. Each document (or research paper) is represented by an input sequence of token embeddings:

$$X = \begin{pmatrix} x_1 \\ x_2 \\ . \\ . \\ . \\ x_T \end{pmatrix} \in \mathbb{R}^{T \times e} \quad (2)$$

where T is the total number of tokens in the document and e is the dimensionality of the embedding space. Each token $x_t$ is an e-dimensional vector. For each category $c_k$, we similarly define a representative sequence of embeddings - these can be drawn from an embedding of the label text itself, or from an external reference corpus, where $T_k$ is the number of tokens (or sub-tokens) used to represent category $c_k$.

$$Y_k = \begin{pmatrix} y_{k,1} \\ y_{k,2} \\ . \\ . \\ . \\ y_{k,T_k} \end{pmatrix} \in \mathbb{R}^{T_k \times e} \quad (3)$$

To derive a single document-level embedding, we introduce an aggregation function $E(\cdot)$ that maps the matrix $X \in \mathbb{R}^{T \times e}$ to a single embedding vector z, as specified in Equation (4):

$$z = E(X) = \sum_{t=1}^{T} \alpha_t \phi(x_t), \text{ with } \sum_{t=1}^{T} \alpha_t = 1 \quad (4)$$

Here, $\phi(x_t)$ is a token-level transform (self-attention), and the nonnegative coefficients $\alpha_t$ form a convex weighting (i.e., they sum to 1). Each category $c_k$ is similarly mapped into a single embedding $z_k$ by:

$$z_k = E'(Y_k) = \sum_{t=1}^{T_k} \beta_{k,t} \psi(y_{k,t}), \text{ with } \sum_{t=1}^{T_k} \beta_{k,t} = 1 \quad (5)$$

Here, the function $\psi(\cdot)$ represents the category-specific adaptation of $\phi(\cdot)$, and the weights $\beta_{k,t}$ form a convex combination. After obtaining the aggregated embeddings z for the document and $z_k$ for the category, we compute a similarity score $s(X, c_k)$, as shown in Equation (6), where $A \in \mathbb{R}^{e \times e}$ is a learned weight matrix, and $b_k \in \mathbb{R}$ is a bias term for category $c_k$. A softmax over all k transforms these raw scores into probabilities, as shown in Equation (7).

$$s(X, c_k) = z^T A\, z_k + b_k, \quad for\ k \in [N] \quad (6)$$

$$\mathcal{P}(c_k \mid X) = \frac{exp(s(X,c_k))}{\sum_{j=1}^{N} exp(s(X,c_j))} \quad (7)$$

Equation (7) generates a distribution $P(c_k \mid X)$ over the set of categories C. In our zero-shot setting, $z_k$ is derived from text descriptions of each category. The model thus classifies a previously unseen document X to the category $c_k$ with the highest $P(c_k \mid X)$. Equation (8) explains that $s(X,c_k)$ can be exponentiated to yield an unnormalized weight for each category. In essence, $\tilde{p}(\cdot)$ corresponds to the numerator in the softmax function from Equation (7).

$$\tilde{p}(c_k \mid X) \triangleq \exp(s(X, c_k)) \quad (8)$$

The total unnormalized weight across all categories is shown in Equation 9. In the earlier notation of Equation (7), Z(X) is the denominator in the softmax function. Equation 10 shows the log form of the probability that is generated after we incorporate the partition function. It parallels the usual form of log(exp(score)/partition).

$$Z(X) \triangleq \sum_{j=1}^{N} \tilde{p}(c_j \mid X) = \sum_{j=1}^{N} \exp(s(X, c_j)) \quad (9)$$

$$\log p(c_k \mid X) = s(X, c_k) - \log Z(X) \quad (10)$$

Equation 11 presents the classification decision strategy: select the category $c_k$ with the highest probability, which is equivalent to selecting the one with the greatest $s(X,c_k)$. Equation 11 thus aligns with the typical single-label assignment strategy. Equation 12 represents an explicit integral representation of this score. The function ω(t) is a weighting function over the continuous index t from 0 to T. Here, ϕ($x_t$) and ψ($c_k$) are embedding transforms, and ⟨·,·⟩ represents a dot product. This formalism generalizes the discrete sum in Equation 4, showing how continuous weighting can define the alignment between a paper's textual stream and a category descriptor.

$$\hat{c}(X) = \arg\max_{c_k \in \mathcal{C}} p(c_k \mid X) = \arg\max_{c_k \in \mathcal{C}} s(X, c_k) \quad (11)$$

$$s(X, c_k) = \int_0^T \omega(t) \langle \phi(x_t), \psi(c_k) \rangle\, dt \quad (12)$$

Equation 13 reconfirms the classification rule by emphasizing a minimization of negative log-likelihood, which is mathematically equivalent to maximizing p(c|X). It is effectively the same outcome as in Equation 11 but expressed from the negative-log-prob angle. The final Equation, Equation 14, confirms that each category's probability is between 0 and 1, and all categories' probabilities sum to 1, consistent with the softmax function from Equation (7). Equation (14) is essentially the standard simplex constraint for a probability distribution.

$$\hat{c}(X) = \arg\min_{c \in \mathcal{C}}\{-\log p(c \mid X)\} \quad (13)$$

$$\forall c_k \in \mathcal{C}: \quad 0 \leq p(c_k \mid X) \leq 1 \quad \text{and} \quad \sum_{k=1}^{N} p(c_k \mid X) = 1 \quad (14)$$

## 3    Results and Discussions

The results obtained from applying the developed zero-shot methodology to the selected set of papers present insightful and informative patterns, as shown in Figures 2 and 3 and tabulated comprehensively in Table 1. These findings offer a meaningful glimpse into the current scholarly discourse surrounding the use of social media data for understanding Long COVID, highlighting both the predominant research themes and areas yet to be sufficiently explored. As shown in Figure 2, the selected papers clustered predominantly around the "Symptom Characterization" category, accounting for 43.6% (17 out of 39) of the total studies analyzed. A careful examination of the papers in this category reveals consistent efforts directed toward cataloging patient-reported symptoms - ranging from fatigue and neurological impairments to various respiratory conditions - as discussed extensively across multiple social media platforms. The considerable prevalence of symptom-focused research is not unexpected; it likely mirrors the medical community's urgent need for detailed symptom profiles of Long COVID, which remain challenging to fully characterize through conventional clinical methodologies alone.

Following symptom-related research, studies categorized under "Community and Support" represent 28.2% (11 papers). This category is notable because it emphasizes an increasingly important dimension of healthcare research: the social and psychological components of disease experiences shared online. These studies explore, in considerable depth, the emotional resilience, coping mechanisms, and collective sensemaking among Long COVID patients, underscoring how digital communities can substantially alter the lived experience of chronic illness. Interestingly, fewer papers were classified under "NLP and Modeling" (17.9%; 7 papers) and "Policy and Advocacy" (10.3%; 4 papers). This smaller proportion hints at these topics' status as emerging rather than fully developed research areas within the Long COVID literature landscape. Papers in "NLP and Modeling" tend to emphasize sophisticated computational techniques, including advanced embedding methodologies, sentiment analysis, and topic modeling. These computational tools reflect recent innovations in natural language processing, demonstrating the potential of such approaches to increase our understanding of public discourse. Conversely, "Policy and Advocacy" papers highlight interactions between online discourse and public health policy - addressing, for instance, how governmental responses, advocacy groups, and patient-led initiatives contribute to shaping Long COVID's public narrative.

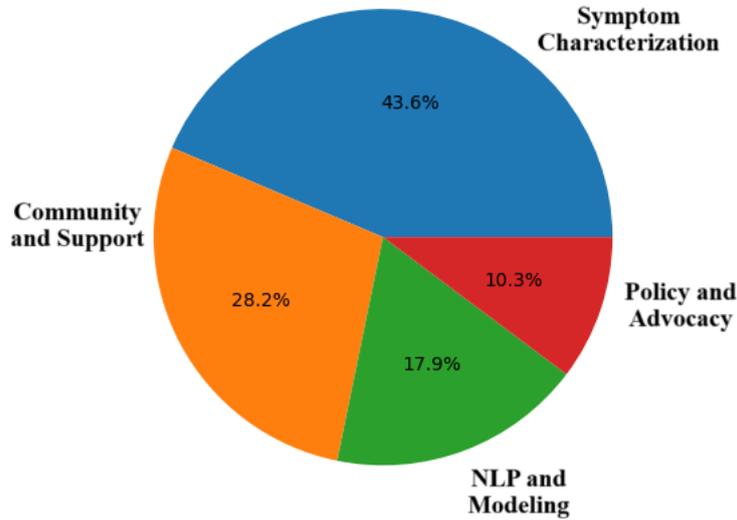

Figure 2. A pie chart-based analysis of the results of the zero-shot classification model

Figure 3 presents a histogram-based analysis of the confidence scores assigned by this model to different papers. These scores ranged from a low of 0.1566 to a high of 0.9928, with an average confidence of 0.7788. The results of the classification for each paper are presented in Table 1. In Table 1, the "Category" column refers to the category to which a paper was classified, and the "Confidence" column refers to the confidence of that classification.

Table 1. Classification of each paper using the zero-shot classification model

| Paper Title | Category | Confidence |
|---|---|---|
| Investigating public perceptions regarding the Long COVID on Twitter using sentiment analysis and topic modeling [67] | Policy and Advocacy | 0.658007 |
| Long Covid – The illness narratives [68] | Community and Support | 0.843067 |
| Support amid uncertainty: Long COVID illness experiences and the role of online communities [69] | Community and Support | 0.980744 |
| The Impact of Polarised Social Media Networking Communications in the #Longcovid Debate between Ideologies and Scientific Facts [70] | Community and Support | 0.7939 |
| An Analysis of Self-reported Longcovid Symptoms on Twitter [71] | Symptom Characterization | 0.841358 |
| Characteristics of Long Covid: findings from a social media survey [72] | Symptom Characterization | 0.939542 |
| Long COVID symptoms from Reddit: Characterizing post-COVID syndrome from patient reports [73] | Symptom Characterization | 0.851446 |
| Characterization of long-term patient-reported symptoms | Symptom | 0.989711 |

Contribution Title (shortened if too long)      11| | | |
|---|---|---|
| of COVID-19: an analysis of social media data [74] | Characterization | |
| Breakthrough Symptomatic COVID-19 Infections Leading to Long Covid: Report from Long Covid Facebook Group Poll [75] | Symptom Characterization | 0.924075 |
| #LongCOVID affects children too: A Twitter analysis of healthcare workers' sentiment and discourse about Long COVID in children and young people in the UK [76] | Community and Support | 0.806065 |
| Using Social Media to Help Understand Long COVID Patient Reported Health Outcomes: A Natural Language Processing Approach [77] | NLP and Modeling | 0.926036 |
| An Interactive Analysis of User-reported Long COVID Symptoms using Twitter Data [78] | Symptom Characterization | 0.508586 |
| Exploring the Emotional and Mental Well-Being of Individuals with Long COVID Through Twitter Analysis [79] | Policy and Advocacy | 0.447378 |
| Understanding the Long Haulers of COVID-19: Mixed Methods Analysis of YouTube Content [80] | NLP and Modeling | 0.794322 |
| The Pulse of Long COVID on Twitter: A Social Network Analysis [81] | Community and Support | 0.526044 |
| Investigating and Analyzing Self-Reporting of Long COVID on Twitter: Findings from Sentiment Analysis [82] | NLP and Modeling | 0.156601 |
| Digital Long-Hauler Lifelines: Understanding How People with Long Covid Build Community on Reddit [83] | Community and Support | 0.985294 |
| The effects of long COVID-19, its severity, and the need for immediate attention: Analysis of clinical trials and Twitter data [84] | NLP and Modeling | 0.438484 |
| Discovering Long COVID Symptom Patterns: Association Rule Mining and Sentiment Analysis in Social Media Tweets [85] | Symptom Characterization | 0.876739 |
| Social Media Mining of Long-COVID Self-Medication Reported by Reddit Users: Feasibility Study to Support Drug Repurposing [86] | NLP and Modeling | 0.836668 |
| The Long COVID experience from a patient's perspective: a clustering analysis of 27,216 Reddit posts [87] | Symptom Characterization | 0.969563 |
| The Role of Social Media in the Experiences of COVID-19 Among Long-Hauler Women: Qualitative Study [88] | Community and Support | 0.955924 |
| State Health Department Communication about Long COVID in the United States on Facebook: Risks, Prevention, and Support [89] | Symptom Characterization | 0.901305 |
| Long COVID at Different Altitudes: A Countrywide Epidemiological Analysis [90] | Symptom Characterization | 0.639446 |
| Understanding the #longCOVID and #longhaulers Conversation on Twitter: Multimethod Study [91] | Community and Support | 0.96493 |
| Identifying Profiles and Symptoms of Patients With Long COVID in France: Data Mining Infodemiology Study Based on Social Media [92] | Symptom Characterization | 0.68115 |
| Using Social Media to Help Understand Patient-Reported Health Outcomes of Post–COVID-19 Condition: Natural Language Processing Approach [93] | NLP and Modeling | 0.868817 |
| Characteristics and impact of Long Covid: Findings from an online survey [94] | Symptom Characterization | 0.908992 |
| Too much focus on your health might be bad for your health: Reddit user's communication style predicts their Long COVID likelihood [95] | Symptom Characterization | 0.650907 |
| An Analysis of Self-reported Long COVID-19 Symptoms on Twitter [96] | Symptom Characterization | 0.852006 |
| Long Covid: Online patient narratives, public health communication and vaccine hesitancy [97] | Community and Support | 0.655784 |
| Understanding The Plight of Covid-19 Long Haulers Through Computational Analysis Of YouTube Content [98] | Community and Support | 0.914525 |
| Twitter Sentiment Analysis of Long COVID Syndrome [99] | Policy and Advocacy | 0.843566 |
| Using Topic Modeling and NLP Tools for Analyzing Long Covid Coverage by French Press and Twitter [100] | NLP and Modeling | 0.60467 |
| A content analysis of the reliability and quality of YouTube videos as a source of information on health-related post-COVID pain [101] | Symptom Characterization | 0.849268 |
| Unlocking the Mysteries of Long COVID in Children and Young People: Insights from a Policy Review and Social Media Analysis in the UK [102] | Policy and Advocacy | 0.938843 |
| Long Haul COVID-19 Videos on YouTube: Implications for Health Communication [103] | Symptom Characterization | 0.34838 |
| The Impact of COVID Vaccination on Symptoms of Long COVID: An International Survey of People with Lived Experience of Long COVID [104] | Symptom Characterization | 0.709309 |
| Social Support and Narrative Sensemaking Online: A Content Analysis of Facebook Posts by COVID-19 Long Haulers [105] | Community and Support | 0.992763 |

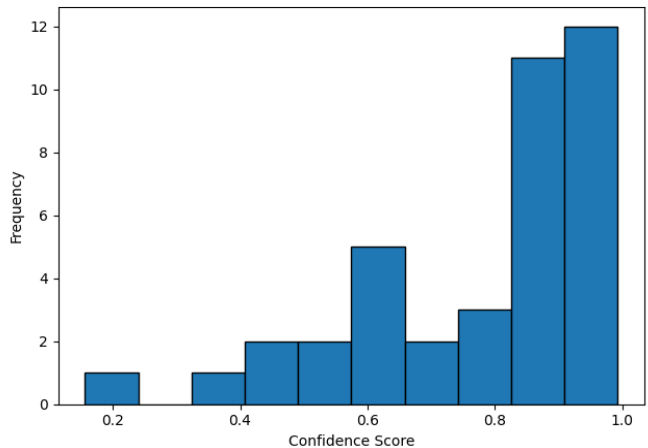

Figure 3. A histogram of confidence scores of the proposed zero-shot classification model

Beyond its immediate utility in the context of classifying studies that focused on the mining and analysis of Long COVID-related content on social media, this model has several applications. The typical process of categorizing research literature relies on manual coding, a labor-intensive task susceptible to individual bias and inconsistency. By leveraging state-of-the-art embedding methods and advanced language models, the proposed methodology eliminates extensive manual effort and reduces subjectivity. Importantly, the zero-shot nature of this technique makes it flexible and capable of rapid adaptation to new or shifting research domains without necessitating specialized, domain-specific training datasets. In public health research and policy, such a model would offer efficient and responsive categorization, especially critical during emerging health crises when rapid information synthesis is essential [108]. Similarly, in sociology and digital humanities, this model could facilitate the rapid exploration of datasets of online discourse, highlighting public attitudes, social concerns, and emerging cultural trends from digital data [109]. The work presented in this paper has a couple of limitations. First, the model used the titles and abstracts of these papers instead of the full texts to ensure computational efficiency and to prevent very high processing times; as a result, potentially meaningful information found exclusively in the main body of texts was not considered, possibly limiting classification accuracy. Second, the model relies on pre-trained language representations and may not fully capture contextual nuances in highly specialized or region-specific vocabulary as expressed on social media.

## 4     Conclusion

Long COVID is becoming a complex health-related problem that requires multiple approaches to comprehensively study the wide range of clinical, long-term effects, and psychosocial aspects associated with it across patients worldwide. Social media has allowed for the collection of patient-generated data in real-time, making it easier to understand the variety of symptoms associated with Long COVID. The work presented in this paper demonstrates that a zero-shot classification strategy - one that relies on textual embeddings rather than manual labels - can effectively map a rapidly growing body of Long COVID literature into distinct thematic categories. By combining sophisticated embedding algorithms and advanced language models, this approach classified research papers in this area into four primary categories: Clinical or Symptom Characterization, Advanced NLP or Computational Methods, Policy, Advocacy or Public Health Communication, and Online Communities and Social Support.

These findings establish that automated classification pipelines can be significantly helpful for categorizing scientific literature in contexts where urgency, complexity, and multidisciplinary perspectives converge. Equally important are the model's confidence estimates, which highlight the reliability of each classification and offer domain experts a means of understanding the strength or ambiguity of a classification. This methodology is expected to open up major opportunities for interdisciplinary research. Furthermore, the process that was followed to develop this model is explained in detail such that researchers in this field can further refine the embedding-based mechanisms to accommodate emerging data sources, while public health practitioners might create customized categories for topics that suddenly gain traction - such as newly identified clinical variants, innovative treatment options, or evolving community advocacy efforts. In summary, the zero-shot classification approach described here highlights a powerful synergy between advanced embedding architectures and real-world requirements: It unburdens researchers from exhaustive manual curation, preserves transparent insight into thematic boundaries, and adapts dynamically to the evolving landscape of the field. Such adaptability is vital for Long COVID research, where new evidence, evolving subtopics, and shifting stakeholder interests demand continuous review.

**Disclosure of Interests.** The authors have no competing interests to declare that are relevant to the content of this article.

## References

1. Ciotti, M., Ciccozzi, M., Terrinoni, A., Jiang, W.-C., Wang, C.-B., Bernardini, S.: The COVID-19 pandemic. Crit. Rev. Clin. Lab. Sci. 57, 365–388 (2020). https://doi.org/10.1080/10408363.2020.1783198.
2. Velavan, T.P., Meyer, C.G.: The COVID-19 epidemic. Trop. Med. Int. Health. 25, 278–280 (2020). https://doi.org/10.1111/tmi.13383.
3. Yesudhas, D., Srivastava, A., Gromiha, M.M.: COVID-19 outbreak: history, mechanism, transmission, structural studies and therapeutics. Infection. 49, 199–213 (2021). https://doi.org/10.1007/s15010-020-01516-2.
4. Allen, D.W.: Covid-19 lockdown cost/benefits: A critical assessment of the literature. Int. J. Econ. Bus. 29, 1–32 (2022). https://doi.org/10.1080/13571516.2021.1976051.
5. Cucinotta, D., Vanelli, M.: WHO Declares COVID-19 a Pandemic. Acta Biomed. Ateneo Parmense. 91, 157–160 (2020). https://doi.org/10.23750/abm.v91i1.9397.
6. COVID-19 cases, https://covid19.who.int/, last accessed 2024/12/23.
7. Raveendran, A.V., Jayadevan, R., Sashidharan, S.: Long COVID: An overview. Diabetes Metab. Syndr. 15, 869–875 (2021). https://doi.org/10.1016/j.dsx.2021.04.007.
8. Altmann, D.M., Whettlock, E.M., Liu, S., Arachchillage, D.J., Boyton, R.J.: The immunology of long COVID. Nat. Rev. Immunol. 23, 618–634 (2023). https://doi.org/10.1038/s41577-023-00904-7.
9. Fernández-de-las-Peñas, C.: Long COVID: current definition. Infection. 50, 285–286 (2022). https://doi.org/10.1007/s15010-021-01696-5.
10. CDC: Long COVID basics, https://www.cdc.gov/covid/long-term-effects/index.html, last accessed 2024/12/23.
11. Committee on Examining the Working Definition for Long COVID, Board on Health Sciences Policy, Board on Global Health, Health and Medicine Division, National Academies of Sciences, Engineering, and Medicine: A long COVID definition: A chronic, systemic disease state with profound consequences, http://dx.doi.org/10.17226/27768, (2024). https://doi.org/10.17226/27768.
12. Aiyegbusi, O.L., et al.: Symptoms, complications and management of long COVID: a review. J. R. Soc. Med. 114, 428–442 (2021). https://doi.org/10.1177/01410768211032850.
13. Subramanian, A., et al.: Symptoms and risk factors for long COVID in non-hospitalized adults. Nat. Med. 28, 1706–1714 (2022). https://doi.org/10.1038/s41591-022-01909-w.

14. Sudre, C.H., et al.: Attributes and predictors of long COVID. Nat. Med. 27, 626–631 (2021). https://doi.org/10.1038/s41591-021-01292-y.
15. Notarte, K.I., et al.: Impact of COVID-19 vaccination on the risk of developing long-COVID and on existing long-COVID symptoms: A systematic review. EClinicalMedicine. 53, 101624 (2022). https://doi.org/10.1016/j.eclinm.2022.101624.
16. Cabrera Martimbianco, A.L., Pacheco, R.L., Bagattini, Â.M., Riera, R.: Frequency, signs and symptoms, and criteria adopted for long COVID-19: A systematic review. Int. J. Clin. Pract. 75, (2021). https://doi.org/10.1111/ijcp.14357.
17. Ayoubkhani, D., et al.: Trajectory of long covid symptoms after covid-19 vaccination: community-based cohort study. BMJ. 377, e069676 (2022). https://doi.org/10.1136/bmj-2021-069676.
18. Davis, H.E., et al.: Characterizing long COVID in an international cohort: 7 months of symptoms and their impact. EClinicalMedicine. 38, 101019 (2021). https://doi.org/10.1016/j.eclinm.2021.101019.
19. Yong, S.J.: Long COVID or post-COVID-19 syndrome: putative pathophysiology, risk factors, and treatments. Infect. Dis. (Lond.). 53, 737–754 (2021). https://doi.org/10.1080/23744235.2021.1924397.
20. Koc, H.C., Xiao, J., Liu, W., Li, Y., Chen, G.: Long COVID and its management. Int. J. Biol. Sci. 18, 4768–4780 (2022). https://doi.org/10.7150/ijbs.75056.
21. Al-Aly, Z., et al.: Long COVID science, research and policy. Nat. Med. 30, 2148–2164 (2024). https://doi.org/10.1038/s41591-024-03173-6.
22. Tana, C., et al.: Long COVID headache. J. Headache Pain. 23, (2022). https://doi.org/10.1186/s10194-022-01450-8.
23. Peluso, M.J., Deeks, S.G.: Mechanisms of long COVID and the path toward therapeutics. Cell. 187, 5500–5529 (2024). https://doi.org/10.1016/j.cell.2024.07.054.
24. Greenhalgh, T., Sivan, M., Perlowski, A., Nikolich, J.Ž.: Long COVID: a clinical update. Lancet. 404, 707–724 (2024). https://doi.org/10.1016/s0140-6736(24)01136-x.
25. Sykes, D.L., et al.: What is long-COVID and how should we manage it? Lung. 199, 113–119 (2021). https://doi.org/10.1007/s00408-021-00423-z.
26. Thakur, N.: Social media mining and analysis: A brief review of recent challenges. Information (Basel). 14, 484 (2023). https://doi.org/10.3390/info14090484.
27. Venegas-Vera, A.V., Colbert, G.B., Lerma, E.V.: Positive and negative impact of social media in the COVID-19 era. Rev. Cardiovasc. Med. 21, (2020). https://doi.org/10.31083/j.rcm.2020.04.195.
28. Thakur, N., Han, C.: An exploratory study of tweets about the SARS-CoV-2 Omicron variant: Insights from sentiment analysis, language interpretation, source tracking, type classification, and embedded URL detection. COVID. 2, 1026–1049 (2022). https://doi.org/10.3390/covid2080076.
29. Thakur, N.: Sentiment analysis and text analysis of the public discourse on Twitter about COVID-19 and MPox. Big Data Cogn. Comput. 7, 116 (2023). https://doi.org/10.3390/bdcc7020116.
30. Hussain, W.: Role of social media in COVID-19 pandemic. Int J Front Sci. 4, (2024). https://doi.org/10.37978/tijfs.v4i2.144.
31. Thakur, N., Cui, S., Khanna, K., Knieling, V., Duggal, Y.N., Shao, M.: Investigation of the gender-specific discourse about online learning during COVID-19 on Twitter using sentiment analysis, subjectivity analysis, and toxicity analysis. Computers. 12, 221 (2023). https://doi.org/10.3390/computers12110221.
32. Southwick, L., Guntuku, S.C., Klinger, E.V., Seltzer, E., McCalpin, H.J., Merchant, R.M.: Characterizing COVID-19 content posted to TikTok: Public sentiment and re-sponse during the first phase of the COVID-19 pandemic. J. Adolesc. Health. 69, 234–241 (2021). https://doi.org/10.1016/j.jadohealth.2021.05.010.
33. Feldkamp, J.: The rise of TikTok: The evolution of a social media platform during COVID-19. In: SpringerBriefs in Information Systems. pp. 73–85. Springer International Publishing, Cham (2021).
34. Piñeiro-Chousa, J., López-Cabarcos, M. Á., Pérez-Pico, A. M., & Caby, J. (2023). The influence of Twitch and sustainability on the stock returns of video game companies: Before and after COVID-19. Journal of business research, 157, 113620.
35. Chae, S. W., & Lee, S. H. (2022). Sharing emotion while spectating video game play: Exploring Twitch users' emotional change after the outbreak of the COVID-19 pandemic. Computers in human behavior, 131, 107211.
36. Wang, W., Wang, Y., Zhang, X., Jia, X., Li, Y., & Dang, S. (2020). Using WeChat, a Chinese social media app, for early detection of the COVID-19 outbreak in December 2019: retrospective study. JMIR mHealth and uHealth, 8(10), e19589.
37. Fan, Z., et al. COVID-19 information dissemination using the WeChat communication index: retrospective analysis study. Journal of Medical Internet Research, 23(7), e28563.
38. Rovetta, A., Bhagavathula, A.S.: Global infodemiology of COVID-19: Analysis of Google web searches and Instagram hashtags. J. Med. Internet Res. 22, e20673 (2020). https://doi.org/10.2196/20673.
39. Malik, A., Khan, M.L., Quan-Haase, A.: Public health agencies outreach through Instagram during the COVID-19 pandemic: Crisis and Emergency Risk Communication perspective. Int. J. Disaster Risk Reduct. 61, 102346 (2021). https://doi.org/10.1016/j.ijdrr.2021.102346.
40. Mejova, Y., Kalimeri, K.: COVID-19 on Facebook ads: Competing agendas around a public health crisis. In: Proceedings of the 3rd ACM SIGCAS Conference on Computing and Sustainable Societies. pp. 22–31. ACM, New York, NY, USA (2020).
41. Perrotta, D., Grow, A., Rampazzo, F., Cimentada, J., Del Fava, E., Gil-Clavel, S., Zagheni, E.: Behaviours and attitudes in response to the COVID-19 pandemic: insights from a cross-national Facebook survey. EPJ Data Sci. 10, 17 (2021). https://doi.org/10.1140/epjds/s13688-021-00270-1.
42. Li, H.O.-Y., Bailey, A., Huynh, D., Chan, J.: YouTube as a source of information on COVID-19: a pandemic of misinformation? BMJ Glob. Health. 5, e002604 (2020). https://doi.org/10.1136/bmjgh-2020-002604.
43. Su, V., Thakur, N.: COVID-19 on YouTube: A data-driven analysis of sentiment, toxicity, and content recommendations. In: 2025 IEEE 15th Annual Computing and Communication Workshop and Conference (CCWC). pp. 00715–00723. IEEE (2025).
44. Veselovsky, V., Anderson, A.: Reddit in the time of COVID. Proceedings of the International AAAI Conference on Web and Social Media. 17, 878–889 (2023). https://doi.org/10.1609/icwsm.v17i1.22196.
45. Hale, B., Alberta, M., Chae, S.W.: Reddit as a Source of COVID-19 Information: A Content Analysis of r/coronavirus during the Early Pandemic. Journal of Communication Technology. 5, 26 (2022).
46. Daglis, T., Tsagarakis, K.P.: A LinkedIn-based analysis of the U.S. dynamic adaptations in healthcare during the COVID-19 pandemic. Healthcare Analytics. 5, 100291 (2024). https://doi.org/10.1016/j.health.2023.100291.
47. Pardim, V.I., Pinochet, L.H.C., Souza, C.A., Viana, A.B.N.: The behavior of young people at the beginning of their career through LinkedIn. RAM Rev. Adm. Mackenzie. 23, eRAMG220064 (2022). https://doi.org/10.1590/1678-6971/eramg220064.en.
48. Rajagumar, P., & Chooi, W. T. (2025). Exploring suicide chatter on Twitter during

COVID-19 lockdown in Malaysia. Discover Public Health, 22(1), 1-12.
49. Thakur, N.: A large-scale dataset of Twitter chatter about online learning during the current COVID-19 Omicron wave. Data (Basel). 7, 109 (2022). https://doi.org/10.3390/data7080109.
50. Hinchey, L., et al.: Clubhouses as essential communities during the COVID-19 pandemic. J. Psychosoc. Rehabil. Ment. Health. 9, 149–157 (2022). https://doi.org/10.1007/s40737-021-00242-8.
51. Junaid, S., Mutschler, C., McShane, K., The Canadian Clubhouse Research Group: The impact of COVID-19 on clubhouse employment programs. Community Ment. Health J. 59, 523–530 (2023). https://doi.org/10.1007/s10597-022-01036-3.
52. Ayob, M.A., Hadi, N.A., Ezad, M., Pahroraji, H.M., Ismail, B., Saaid, M.N.F.: Promoting 'Discord' as a platform for learning engagement during Covid-19 pandemic. Asian J. Univ. Educ. 18, 663–673 (2022). https://doi.org/10.24191/ajue.v18i3.18953.
53. Ardiyansah, T.Y., Batubara, R.W., Auliya, P.K.: Using discord to facilitate students in teaching learning process during COVID-19 outbreak. Journal of English Teaching, Literature, and Applied Linguistics. 5, 76 (2021). https://doi.org/10.30587/jetlal.v5i1.2528.
54. Yang, Q., et al.: Online communication shifts in the midst of the Covid-19 pandemic: A case study on Snapchat. Proceedings of the International AAAI Conference on Web and Social Media. 15, 830–840 (2021). https://doi.org/10.1609/icwsm.v15i1.18107.
55. Spieler, B., et al.: Diagnosis in a snap: a pilot study using Snapchat in radiologic didactics. Emerg. Radiol. 28, 93–102 (2021). https://doi.org/10.1007/s10140-020-01825-x.
56. Thakur, N., Han, C.Y.: A multimodal approach for early detection of cognitive impairment from tweets. In: Human Interaction, Emerging Technologies and Future Systems V. pp. 11–19. Springer International Publishing, Cham (2022).
57. Midgley, C., Lockwood, P., Thai, S.: Can the social network bridge social distancing? Social media use during the COVID-19 pandemic. Psychology of Popular Media. 13, 44–54 (2024). https://doi.org/10.1037/ppm0000437.
58. Thakur, N.: MonkeyPox2022Tweets: A large-scale Twitter dataset on the 2022 Monkeypox outbreak, findings from analysis of Tweets, and open research questions. Infect. Dis. Rep. 14, 855–883 (2022). https://doi.org/10.3390/idr14060087.
59. Perego, E.: #LongCovid, https://twitter.com/elisaperego78/status/1263172084055838721?s=20, last accessed 2024/12/23.
60. Thakur, N., Cho, H., Cheng, H., Lee, H.: Analysis of user diversity-based patterns of public discourse on twitter about mental health in the context of online learning during COVID-19. In: Lecture Notes in Computer Science. pp. 367–389. Springer Nature Switzerland, Cham (2023).
61. Schillinger, D., Chittamuru, D., Ramírez, A.S.: From "infodemics" to health promotion: A novel framework for the role of social media in public health. Am. J. Public Health. 110, 1393–1396 (2020). https://doi.org/10.2105/ajph.2020.305746.
62. Utari, U., Wulandari, Y., Colby, C., Crespi, C.: Political participation of the Millennial generation in general elections: The influence of education, social media, and economic factors. jiph. 12, 183–198 (2023). https://doi.org/10.35335/jiph.v12i3.10.
63. Injadat, M., Salo, F., & Nassif, A. B. (2016). Data mining techniques in social media: A survey. Neurocomputing, 214, 654-670.
64. Pourpanah, F., Abdar, M., Luo, Y., Zhou, X., Wang, R., Lim, C.P., Wang, X.-Z., Wu, Q.M.J.: A review of generalized zero-shot learning methods. IEEE Trans. Pattern Anal. Mach. Intell. 45, 1–20 (2022). https://doi.org/10.1109/tpami.2022.3191696.
65. Romera-Paredes, B., Torr, P.H.S.: An embarrassingly simple approach to zero-shot

learning. ICML. 37, 2152–2161 (07--July 09 2015). https://doi.org/10.1007/978-3-319-50077-5_2.
66. Wang, W., Zheng, V.W., Yu, H., Miao, C.: A survey of zero-shot learning: Settings, methods, and applications. ACM Trans. Intell. Syst. Technol. 10, 1–37 (2019). https://doi.org/10.1145/3293318.
67. Fu, Y.: Investigating public perceptions regarding the Long COVID on Twitter using sentiment analysis and topic modeling. Med. Data Min. (2022). https://doi.org/10.53388/mdm20220520024.
68. Rushforth, A., Ladds, E., Wieringa, S., Taylor, S., Husain, L., Greenhalgh, T.: Long Covid – The illness narratives. Soc. Sci. Med. 286, 114326 (2021). https://doi.org/10.1016/j.socscimed.2021.114326.
69. Russell, D., Spence, N.J., Chase, J.-A.D., Schwartz, T., Tumminello, C.M., Bouldin, E.: Support amid uncertainty: Long COVID illness experiences and the role of online communities. SSM Qual. Res. Health. 2, 100177 (2022). https://doi.org/10.1016/j.ssmqr.2022.100177.
70. Meledandri, F.: The impact of polarised social media networking communications in the #longcovid debate between ideologies and scientific facts, http://dx.doi.org/10.13136/2281-4582/2024.I23.1450, (2024).
71. Singh, S.M., Reddy, C.: An analysis of self-reported longcovid symptoms on twitter, http://dx.doi.org/10.1101/2020.08.14.20175059, (2020).
72. Ziauddeen, N., Gurdasani, D., O'Hara, M.E., Hastie, C., Roderick, P., Yao, G., Alwan, N.A.: Characteristics of Long Covid: findings from a social media survey, http://dx.doi.org/10.1101/2021.03.21.21253968, (2021).
73. Sarker, A., Ge, Y.: Long COVID symptoms from Reddit: Characterizing post-COVID syndrome from patient reports, http://dx.doi.org/10.1101/2021.06.15.21259004, (2021).
74. Banda, J.M., et al.: Characterization of long-term patient-reported symptoms of COVID-19: an analysis of social media data, http://dx.doi.org/10.1101/2021.07.13.21260449, (2021).
75. Massey, D., Berrent, D., Krumholz, H.: Breakthrough symptomatic COVID-19 infections leading to Long Covid: Report from Long Covid Facebook group poll, http://dx.doi.org/10.1101/2021.07.23.21261030, (2021).
76. Martin, S., Chepo, M., Déom, N., Khalid, A.F., Vindrola-Padros, C.: "#LongCOVID affects children too": A Twitter analysis of healthcare workers' sentiment and discourse about Long COVID in children and young people in the UK, http://dx.doi.org/10.1101/2022.07.20.22277865, (2022).
77. Dolatabadi, E., et al.: Using social media to help understand long COVID patient reported health outcomes: A natural language processing approach, http://dx.doi.org/10.1101/2022.12.14.22283419, (2022).
78. Miao, L., Last, M., Litvak, M.: An interactive analysis of user-reported long COVID symptoms using twitter data. In: Hruschka, E., Mitchell, T., Mladenic, D., Grobelnik, M., and Bhutani, N. (eds.) Proceedings of the 2nd Workshop on Deriving Insights from User-Generated Text. pp. 10–19. Association for Computational Linguistics, Stroudsburg, PA, USA (2022).
79. Guocheng, F., Huaiyu, C., Wei, Q.: Exploring the emotional and mental well-being of individuals with Long COVID through twitter analysis, http://arxiv.org/abs/2307.07558, (2023).
80. Jordan, A., Park, A.: Understanding the long haulers of COVID-19: Mixed methods analysis of YouTube content. JMIR AI. 3, e54501 (2024). https://doi.org/10.2196/54501.

81. Kusuma, I.Y., Suherman, S.: The pulse of long COVID on Twitter: A social network analysis. Arch. Iran. Med. 27, 36–43 (2024). https://doi.org/10.34172/aim.2024.06.
82. Thakur, N.: Investigating and analyzing self-reporting of Long COVID on Twitter: Findings from sentiment analysis. Appl. Syst. Innov. 6, 92 (2023). https://doi.org/10.3390/asi6050092.
83. Digital Long-Hauler Lifelines: Understanding How People With Long Covid Build Community On Reddit, https://www.researchgate.net/publication/385720439_Digital_Long-Hauler_Lifelines_Understanding_How_People_with_Long_Covid_Build_Community_on_Reddit, last accessed 2024/12/24.
84. Bhattacharyya, A., Seth, A., Rai, S.: The effects of long COVID-19, its severity, and the need for immediate attention: Analysis of clinical trials and Twitter data. Front. Big Data. 5, (2022). https://doi.org/10.3389/fdata.2022.1051386.
85. Matharaarachchi, S., Domaratzki, M., Katz, A., Muthukumarana, S.: Discovering long COVID symptom patterns: Association rule mining and sentiment analysis in social media tweets. JMIR Form. Res. 6, e37984 (2022). https://doi.org/10.2196/37984.
86. Koss, J., Bohnet-Joschko, S.: Social media mining of long-COVID self-medication reported by Reddit users: Feasibility study to support drug repurposing. JMIR Form. Res. 6, e39582 (2022). https://doi.org/10.2196/39582.
87. Ayadi, H., Bour, C., Fischer, A., Ghoniem, M., Fagherazzi, G.: The Long COVID experience from a patient's perspective: a clustering analysis of 27,216 Reddit posts. Front. Public Health. 11, (2023). https://doi.org/10.3389/fpubh.2023.1227807.
88. Garrett, C., Aghaei, A., Aggarwal, A., Qiao, S.: The role of social media in the experiences of COVID-19 among long-hauler women: Qualitative study. JMIR Hum. Factors. 11, e50443 (2024). https://doi.org/10.2196/50443.
89. Laestadius, L.I., Guidry, J.P.D., Bishop, A., Campos-Castillo, C.: State health department communication about long COVID in the United States on Facebook: Risks, prevention, and support. Int. J. Environ. Res. Public Health. 19, 5973 (2022). https://doi.org/10.3390/ijerph19105973.
90. Izquierdo-Condoy, J.S., et al.: Long COVID at different altitudes: A countrywide epidemiological analysis. Int. J. Environ. Res. Public Health. 19, 14673 (2022). https://doi.org/10.3390/ijerph192214673.
91. Santarossa, S., et al.: Understanding the #longCOVID and #longhaulers conversation on Twitter: Multimethod study. JMIR Infodemiology. 2, e31259 (2022). https://doi.org/10.2196/31259.
92. Déguilhem, A., et al.: Identifying profiles and symptoms of patients with long COVID in France: Data mining infodemiology study based on social media. JMIR Infodemiology. 2, e39849 (2022). https://doi.org/10.2196/39849.
93. Dolatabadi, E., et al.: Using social media to help understand patient-reported health outcomes of post–COVID-19 condition: Natural language processing approach. J. Med. Internet Res. 25, e45767 (2023). https://doi.org/10.2196/45767.
94. Ziauddeen, N., Gurdasani, D., O'Hara, M.E., Hastie, C., Roderick, P., Yao, G., Alwan, N.A.: Characteristics and impact of Long Covid: Findings from an online survey. PLoS One. 17, e0264331 (2022). https://doi.org/10.1371/journal.pone.0264331.
95. Segneri, L., Babina, N., Hammerschmidt, T., Fronzetti Colladon, A., Gloor, P.A.: Too much focus on your health might be bad for your health: Reddit user's communication style predicts their Long COVID likelihood. PLoS One. 19, e0308340 (2024). https://doi.org/10.1371/journal.pone.0308340.
96. Singh, S.M., Reddy, S.C., Kathiravan, S.: An analysis of self-reported long COVID-19

symptoms on twitter. J. Postgrad. Med. Educ. Res. 57, 79–81 (2023). https://doi.org/10.5005/jp-journals-10028-1616.
97. Miyake, E., Martin, S.: Long Covid: Online patient narratives, public health communication and vaccine hesitancy. Digit. Health. 7, (2021). https://doi.org/10.1177/20552076211059649.
98. Jordan, A.A.D.: Understanding the plight of covid-19 long haulers through computational analysis of YouTube content, (2022).
99. Awoyemi, T., Ebili, U., Olusanya, A., Ogunniyi, K.E., Adejumo, A.V.: Twitter sentiment analysis of long COVID syndrome. Cureus. 14, e25901 (2022). https://doi.org/10.7759/cureus.25901.
100. Minel, B.J.A.: Using topic modeling and NLP tools for analyzing long Covid coverage by French press and Twitter. In: Nagar et al, A. (ed.) Intelligent Sustainable Systems, Lecture Notes in Networks and Systems 817. Springer Nature Singapore, Singapore (2024). https://doi.org/10.1007/978-981-99-7886-1_15.
101. Ozduran, E., Büyükçoban, S.: A content analysis of the reliability and quality of Youtube videos as a source of information on health-related post-COVID pain. PeerJ. 10, e14089 (2022). https://doi.org/10.7717/peerj.14089.
102. Déom, N., et al.: Unlocking the mysteries of long COVID in children and young people: Insights from a policy review and social media analysis in the UK, https://osf.io/preprints/f48yg/, (2023). https://doi.org/10.31219/osf.io/f48yg.
103. Jacques, E.T., Basch, C.H., Park, E., Kollia, B., Barry, E.: Long haul COVID-19 videos on YouTube: Implications for health communication. J. Community Health. 47, 610–615 (2022). https://doi.org/10.1007/s10900-022-01086-4.
104. Strain, W.D., et al.: The impact of COVID vaccination on symptoms of long COVID: An international survey of people with lived experience of long COVID. Vaccines (Basel). 10, 652 (2022). https://doi.org/10.3390/vaccines10050652.
105. Wongtavavimarn, K.: Social support and narrative sensemaking online: A content analysis of Facebook posts by COVID-19 long haulers, https://uh-ir.tdl.org/bitstream/handle/10657/10745/WONGTAVAVIMARN-THESIS-2022.pdf?sequence=1, last accessed 2024/12/24.
106. Helmy, Y.A., et al.: The COVID-19 pandemic: A comprehensive review of taxonomy, genetics, epidemiology, diagnosis, treatment, and control. J. Clin. Med. 9, 1225 (2020). https://doi.org/10.3390/jcm9041225.
107. Gasser, U., Ienca, M., Scheibner, J., Sleigh, J., Vayena, E.: Digital tools against COVID-19: taxonomy, ethical challenges, and navigation aid. Lancet Digit. Health. 2, e425–e434 (2020). https://doi.org/10.1016/s2589-7500(20)30137-0.
108. Palagyi, A., Marais, B. J., Abimbola, S., Topp, S. M., McBryde, E. S., & Negin, J. (2019). Health system preparedness for emerging infectious diseases: a synthesis of the literature. *Global Public Health*, *14*(12), 1847-1868.
109. Longhi, J. (2020). Proposals for a discourse analysis practice integrated into digital humanities: theoretical issues, practical applications, and methodological consequences. *Languages*, *5*(1), 5.